\documentclass[a4paper,10pt]{article}
\usepackage{graphicx}
\usepackage{comment}
\usepackage{amssymb}
\usepackage{amsmath}
\usepackage{nicefrac}
\usepackage{graphicx}
\usepackage{dcolumn}
\usepackage{bm}
\usepackage{comment}
\usepackage{multirow}
\usepackage{cite}

\begin{document}

\newcommand{\bin}[2]{\left(\begin{array}{c}\!#1\!\\\!#2\!\end{array}\right)}

\huge

\begin{center}
On the vacuum-polarization Uehling potential for a Fermi charge distribution
\end{center}

\vspace{0.5cm}

\large

\begin{center}
Jean-Christophe Pain\footnote{jean-christophe.pain@cea.fr}
\end{center}

\normalsize

\begin{center}
\it CEA, DAM, DIF, F-91297 Arpajon, France
\end{center}

\vspace{0.5cm}

\begin{abstract}
We present analytical formulas for the vacuum-polarization Uehling potential in the case where the finite size of the nucleus is modeled by a Fermi charge distribution. Using a Sommerfeld-type development, the potential is expressed in terms of multiple derivatives of a particular integral. The latter and its derivatives can be evaluated exactly in terms of Bickley-Naylor functions, whose connection to the Uehling potential was already pointed out in the pure Coulomb case, and of usual Bessel functions of the second kind. The cusp and asymptotic expressions for the Uehling potential with a Fermi charge distribution are also provided. Analytical results for the higher-order-contribution K\"all\`en-Sabry potential are given.
\end{abstract}
\section{Introduction}
\label{intro}
Vacuum polarization in light atoms and ions is an important topic of Quantum Electrodynamics (QED) \cite{Akhiezer81,Greiner10}. In 1935, Uehling proposed a formula for the interaction potential between two point-like electric charges which contains an additional term responsible for the electric polarization of the vacuum \cite{Uehling35}. The evaluation of the vacuum-polarization potential of a point charge moving in the Coulomb field of a nucleus is a hard task. The vacuum-polarization correction for an electron in a nuclear Coulomb field can be described, up to the first order in $\alpha$, by a correction to the Coulomb potential:

\begin{eqnarray}
V(r)=-\frac{Ze^2}{4\pi\epsilon_0 r}\left[1+\frac{2}{3\pi}\frac{e^2}{4\pi\epsilon_0\hbar c}\int_1^{\infty}\sqrt{t^2-1}\left(\frac{1}{t^2}+\frac{1}{2t^4}\right)e^{-2mctr/\hbar}dt\right],\nonumber\\
\end{eqnarray}

\noindent where $Z$ is the nuclear charge, $e$ the electron charge, $\epsilon_0$ the dielectric constant, $\hbar$ the reduced Planck constant, and $c$ the speed of light. We can write, in atomic units ($m=\hbar=e=1$) and setting $4\pi\epsilon_0$=1:

\begin{equation}
V(r)=-\frac{Z}{r}+\delta V(r),
\end{equation}

\noindent where $\delta V(r)$ reads

\begin{equation}\label{eq1}
\delta V(r)=-\frac{2\alpha Z}{3\pi r}\int_1^{\infty}\sqrt{t^2-1}\left(\frac{1}{t^2}+\frac{1}{2t^4}\right)e^{-2ctr}dt,
\end{equation} 

\noindent referred to as the Uehling potential \cite{Uehling35}. We keep $\alpha$ and $c$ in the same equation, although in atomic units (which will be used throughout the paper), one has $\alpha=1/c$. Formula (\ref{eq1}) was obtained from Eq. (44) of Wichmann and Kroll \cite{Wichmann56}, using the transformation $t=\sqrt{y^2+1}$. The integral is usually evaluated numerically, but it is worth mentioning that Pyykk\"o et al. \cite{Pyykko98,Pyykko03} derived a two-parameter fitting expression:

\begin{eqnarray}
\delta V(r)=-\frac{\alpha Z}{r}\left[\vphantom{\left(\frac{r}{\alpha}\right)^{0.5}+\left(\frac{r}{\alpha}\right)^{1.5}}e^{-d_1r^ 2}c_1\left(\ln\left(\frac{r}{\alpha}\right)-c_2\right)+\frac{(1-e^{-d_1r^2})}{c_3}\frac{e^{-2r/\alpha}}{d_2\left(\frac{r}{\alpha}\right)^{0.5}+\left(\frac{r}{\alpha}\right)^{1.5}}\right],
\end{eqnarray}

\noindent with $c_1=2/(2\pi)$, $c_2=5/6+\gamma_E$, $\gamma_E$ being the Euler-Mascheroni constant \cite{Abramowitz72}, $c_3=4\sqrt{\pi}$, $d_1=0.678~10^7$ and $d_2$=1.4302 (the formula (A1) of Ref. \cite{Pyykko98} contains an error which was corrected in Ref. \cite{Pyykko03}: $\alpha/c_3$ must be replaced by $1/c_3$). An exact expression, in terms of Sine and Cosine integral functions, was recently obtained by Mez\H{o} \cite{Mezo16}:

\begin{eqnarray}\label{mez}
\delta V(r)=-\frac{4\alpha Z}{\pi r}\int_0^1x(1-x)\left[\mathrm{Shi}\left(\frac{cr}{\sqrt{x(1-x)}}\right)-\mathrm{Chi}\left(\frac{cr}{\sqrt{x(1-x)}}\right)\right]dx,
\end{eqnarray}

\noindent where

\begin{equation}
\mathrm{Shi}(z)=\int_0^z\frac{\sinh(t)}{t}dt
\end{equation}

\noindent is the hyperbolic sine integral and

\begin{equation}
\mathrm{Chi}(z)=\gamma_E+\ln(z)+\int_0^ z\frac{\cosh(t)-1}{t}dt
\end{equation}

\noindent the hyperbolic cosine integral. The limit of $\delta V(r)$ for $cr\ll 1$ was derived by Berestetskii, Lifshitz and Pitaevskii in Ref. \cite{{Berestetskii82}}, but the calculation is rather tedious. We show that Eq. (\ref{mez}) enables one to obtain the result immediately. Noticing that only $\mathrm{Chi}(z)$ will contribute and that

\begin{equation}
\mathrm{Chi}(z)\approx\gamma_E+\ln(z) \;\;\;\; \mathrm{when} \;\;\;\; z\ll 1,
\end{equation}

\noindent we get, since

\begin{equation}
\int_0^1x(1-x)dx=\frac{1}{6}
\end{equation}

\noindent and 
 
\begin{equation}
\int_0^1x(1-x)\ln\left[\frac{cr}{\sqrt{x(1-x})}\right]dx=\frac{5+6\ln(cr)}{36},
\end{equation}
 
\noindent the asymptotic form

\begin{equation}\label{as1}
\delta V(r)\approx-\frac{2\alpha Z}{3\pi r}\left[-\gamma_E-\frac{5}{6}+\ln\left(\frac{1}{cr}\right)\right] \;\;\;\; \mathrm{when} \;\;\;\; r\ll \frac{1}{c}.
\end{equation}

Several methods have been proposed to calculate the integral of Eq. (\ref{eq1}): for instance, Huang \cite{Huang76} derived series expansions which converge for all values of $r$, whereas Fulleton and Rinker \cite{Fullerton76} have found rational approximations. Klarsfeld \cite{Klarsfeld77}, however, expressed the Uehling potential simply in terms of Bessel functions and their integrals, thus generalizing a formula by Pauli and Rose \cite{Pauli36} showing, that way, that on the contrary to what is mentioned in numerous textbooks on quantum mechanics, an analytical expression does exist. Such expression, which involves Bickley-Naylor functions \cite{Bickley35a,Bickley35b,Blair78}, was rediscovered by Frolov and Wardlaw \cite{Frolov12,Frolov14} in a different way. The main problem is reduced to the evaluation of Bessel functions and their integrals. 

However, no such formula is available for an arbitrary charge distribution. Hnizdo proposed to perform calculations in the reciprocal space, using Fourier transforms \cite{Hnizdo93}. In Ref. \cite{Frolov13}, Frolov derived the expression of the lowest-order correction to the vacuum polarization which contains the electron-density function $\rho(x)$. Ginges and Berengut \cite{Ginges16} derived an exact expression assuming a step-function density (homogeneous distribution) for the nucleus. The purpose of the present article is not to provide a numerically efficient method to compute the Uehling potential for a Fermi charge distribution. The most powerful method to do so is probably the rational approximation derived by Fullerton and Rinker. Indeed, the latter method provides nine-digit accuracy with a minimal amount of computation. Thus, in the case of a Fermi charge distribution, the most natural solution is to use this rational approximation, and to evaluate the integral by standard numerical integration methods. Our goal here is to exhibit exact mathematical expressions and relations which may be of interest or raise new ideas in the field. Pointing out connections with special functions may bring new insights, through the use of recursion relations, derivation of asymptotic expressions, connection with other physical problems or extension to other types of densities.  In Sec. \ref{sec1}, we propose an analytical formula for the Uehling potential in case of a Fermi charge distribution (sometimes called Woods-Saxon distribution \cite{Woods54}), which is the most commonly used. The result, obtained thanks to a Sommerfeld-type development, involves multiple derivatives of an integral, which resembles the integral involved in the pure Coulomb case, but with a higher power of the integration variable in the denominator of the integrand. In Sec. \ref{sec2}, we mention that the multiple derivative can be estimated using the expansions of McKee and Glauber \emph{et al.}. In Sec. \ref{sec3}, an exact expression of these derivatives is obtained, involving Bickley-Naylor functions \cite{Bickley35a,Bickley35b,Blair78}, whose connection to the Uehling potential was emphasized by Frolov and Wardlaw in the pure Coulomb case \cite{Frolov12}, and usual Bessel functions of the second kind. In Sec. \ref{sec3}, the asymptotic expressions for $cr\ll 1$ and $cr\gg 1$ are discussed and analytical results for the higher-order-contribution K\"all\`en-Sabry potential are given in Sec. \ref{sec4}.

%The obtained closed form can in addition be of numerical interest due to the recursion relations satisfied by Bickley-Naylor and Bessel functions. 

\section{Exact expansion of the potential for a Fermi density}\label{sec1}

For a charge distribution $\rho(x)$ normalized to

\begin{equation}\label{norm}
\int d^3x\rho(x)=4\pi\int_0^{\infty}x^2\rho(x)dx=Z,
\end{equation}

\noindent the Uehling potential $V(r)$ may be generalized to \cite{Johnson}

\begin{eqnarray}
\delta V(r)=-\frac{2\alpha Z}{3\pi r}\int d^3x\rho(x)\int_1^{\infty}dt\sqrt{t^2-1}\left(\frac{1}{t^2}+\frac{1}{2t^4}\right)\frac{e^{-2ctR}}{R},
\end{eqnarray}

\noindent with $R=|\vec{r}-\vec{x}|=\sqrt{r^2-2rx\cos(\theta)+x^2}$. The integral over $\theta$ and $\phi$, the angles of $\vec{x}$ in spherical coordinates, can be expressed as

\begin{eqnarray}
J(x,r)=\int_0^{2\pi}\int_0^{\pi}\frac{e^{-\lambda R}}{R}d\Omega=2\pi\int_{-1}^1\frac{e^{-\lambda\sqrt{r^2-2rx\mu+x^2}}}{\sqrt{r^2-2rx\mu+x^2}}d\mu.
\end{eqnarray}

\noindent Using $R$ as independent variable, we find

\begin{eqnarray}
J(x,r)=-\frac{2\pi}{rx}\int_{r+x}^{|r-x|}e^{-\lambda R}dR=\frac{2\pi}{\lambda r x}\left[e^{-\lambda|r-x|}-e^{-\lambda (r+x)}\right].
\end{eqnarray}

\noindent We may therefore express $\delta V(r)$ as

\begin{eqnarray}\label{jo}
\delta V(r)=-\frac{2\alpha^2}{3r}\int_0^{\infty}dxx\rho(x)\int_1^{\infty}\sqrt{t^2-1}\left(\frac{1}{t^3}+\frac{1}{2t^5}\right)\left(e^{-2ct|r-x|}-e^{-2ct(r+x)}\right)dt,
\end{eqnarray}

\noindent which is the form given by Fullerton and Rinker \cite{Fullerton76}. Setting, $z=2c(r+x)$, let us define the integral

\begin{equation}\label{defg}
g(z)=\int_1^{\infty}\sqrt{t^2-1}\left(\frac{1}{t^3}+\frac{1}{2t^5}\right)e^{-zt}dt
\end{equation}

\noindent and consider the case of a Fermi-like distribution 

\begin{equation}
\rho(x)=\rho_0f(x)=\frac{\rho_0}{1+e^{\left[(x-\xi)/a\right]}}.
\end{equation}

\noindent The surface thickness is equal to $a=t/(4\ln 3)$ with $t$=2.3 fm \cite{Fricke71} and $\xi=2.2677$ 10$^{-5}$ $a_0$, $a_0$ being the Bohr radius. For recent reviews on the finite nuclear charge distributions, see Ref. \cite{Andrae00}. Several methods have been developed to evaluate the integrals involving Fermi functions. They often rely on a particular representation of the Fermi distribution, such as the Matsubara expansion \cite{Mahan81} or an infinite sum of contour integrals in the complex energy plane \cite{Goedecker93}, etc. In order to calculate the integral 

\begin{equation}
\int_0^{\infty}dx\rho(x)xg(2c(r+x))=\rho_0\int_0^{\infty}dxf(x)xg(2c(r+x)),
\end{equation} 

\noindent it is interesting to resort to the following exact development \cite{Grypeos98} (called here abusively ``Sommerfeld-like'' expansion as a reference to a similar expression introduced by Sommerfeld in solid-state physics \cite{Ashcroft76}):

\begin{eqnarray}\label{sum}
\int_0^{\infty}f(y)H(y)dy=\int_0^{\xi}H(y)dy+\sum_{n=0}^{\infty}a^{2n+2}\left(2-\frac{1}{2^{2n}}\right)\zeta(2n+2)\frac{d^{2n+1}H}{dy^{2n+1}}(\xi)+\mathcal{R},
\end{eqnarray}

\noindent where $\zeta$ represents the Riemann zeta function

\begin{equation}
\zeta(x)=\sum_{n=1}^{\infty}\frac{1}{n^x}
\end{equation}

\noindent which is related to the Bernoulli numbers $B_n$ by 

\begin{equation}
\zeta(2n)=2^{2n-1}\frac{\pi^ {2n}}{(2n)!}B_n
\end{equation}

\noindent and

\begin{equation}
\zeta(2)=\frac{\pi^2}{6}\;\;\;\;\mathrm{and}\;\;\;\;\zeta(4)=\frac{\pi^4}{90},
\end{equation}

\noindent the residual term being equal to

\begin{equation}
\mathcal{R}=\sum_{n=1}^{\infty}(-1)^ne^{-n\xi/a}\int_0^{\infty}H(y)e^{-ny/a}dy.
\end{equation}

\noindent The normalization condition (\ref{norm}) gives

\begin{equation}
\rho_0=\frac{3Z}{4\pi \xi^3\mathcal{N}},
\end{equation}

\noindent with

\begin{equation}
\mathcal{N}=1+\pi^2\frac{a^2}{\xi^2}+6\frac{a^3}{\xi^3}\sum_{n=1}^{\infty}\frac{(-1)^{n-1}}{n^3}e^{-n\xi/a}.
\end{equation}

Inspection of Eq. (\ref{sum}) shows that the problem boils down to the calculation of multiple derivatives of $g(z)$.

\section{Exact expressions using Bickley-Naylor functions}\label{sec2}

\subsection{Expansion in terms of the exponential integral function}

Considering the general class of functions:

\begin{equation}
\chi_n(z)=\int_1^{\infty}\frac{1}{t^n}\left(1+\frac{1}{2t^2}\right)\left(1-\frac{1}{t^2}\right)^{1/2}e^{-zt}dt.
\end{equation}

\noindent it is possible to write \cite{Huang76}

\begin{equation}
\chi_2(z)=f_a(z)E_1(z)+f_b(z)e^{-z}
\end{equation}

\noindent where $f_a$ and $f_b$ are entire functions of $z$ and

\begin{equation}\label{ei}
E_1(z)=\int_1^{\infty}\frac{e^{-zt}}{t}dt
\end{equation}

\noindent is the exponential integral. We have $g(z)=\chi_2(z)$. The functions $f_a$ and $f_b$ can be expanded in power series

\begin{equation}
f_a(z)=\sum_{k=0}^{\infty}C_kz^{2k+1}
\end{equation}

\noindent and

\begin{equation}
f_b(z)=\sum_{k=0}^{\infty}D_kz^{2k+1}
\end{equation}

\noindent where the coefficients $C_k$ and $D_k$ have been obtained by McKee \cite{Mckee69} following Glauber \emph{et al.} \cite{Glauber60}. Then $d^ng(z)/dz^n$ is simple to obtain using Leibniz formula for the derivative of a product. It involves multiple derivative of the exponential integral function of Eq. (\ref{ei}). One has:

\begin{eqnarray}
\frac{d^n}{dx^n}E_1(z)=(-1)^ne^{-z}\sum_{k=0}^{n-1}\bin{n-1}{k}\frac{k!}{z^{k+1}}=(-1)^ne^{-z}\sum_{k=0}^{n-1}\frac{(n-1)!}{(n-1-k)!}\frac{1}{z^{k+1}}=\left(-\frac{1}{z}\right)^n\Gamma\left(n,z\right),
\end{eqnarray}

\noindent where

\begin{equation}
\Gamma\left(n,z\right)=\int_z^{\infty}t^ne^{-t}dt
\end{equation}

\noindent is the incomplete Gamma function. Roesel calculated the function $g(z)$ using an expansion in terms of Chebyshev polynomials which represent rapidly converging series \cite{Roesel77,Luke}. In contrast to rational approximation, such an approach leads to an expansion where the accuracy is only determined by the number of terms taken into account in the series. 

\subsection{Expansion in terms of Bickley-Naylor functions}

The above formulas are definitely useful, but in the following we show that it is possible to derive an exact formulation of $g(z)$ in terms of Bickley-Naylor functions, formulation that may be of interest for obtaining exact expressions of the Uehling potential in case of a Fermi charge distribution. Making the change of variables $t\rightarrow \cosh(u)$ in Eq. (\ref{defg}), we get

\begin{eqnarray}\label{g1}
g(z)=\int_0^1e^{-z\cosh(u)}\left(\frac{1}{\cosh(u)}-\frac{1}{2\cosh^3(u)}-\frac{1}{2\cosh^5(u)}\right)du=-Ki_1(z)+\frac{1}{2}Ki_3(z)+\frac{1}{2}Ki_5(z)
\end{eqnarray}

\noindent where $Ki_n(z)$ is the Bickley-Naylor function defined as

\begin{equation}
Ki_n(z)=\int_0^{\infty}\frac{e^{-z\cosh(t)}}{\cosh^n(t)}dt
\end{equation}

\noindent with $Ki_0(z)=K_0(z)$ the modified Bessel function of zeroth order. The Bickley-Naylor functions satisfy the following differentiation and integration rules

\begin{equation}
\frac{d}{dz}Ki_{n+1}(z)=-Ki_n(z)
\end{equation}

\noindent and

\begin{equation}
Ki_{n+1}(z)=\int_z^{\infty}Ki_n(y)dy.
\end{equation}

\noindent They also obey the following recursion relation

\begin{eqnarray}\label{recur}
(n-1)Ki_n(z)=(n-2)Ki_{n-2}(z)+z\left[Ki_{n-3}(z)-Ki_{n-1}(z)\right]
\end{eqnarray}

\noindent and follow the asymptotic form \cite{Abramowitz72}

\begin{equation}
Ki_n(z)\approx\sqrt{\frac{\pi}{2z}}e^{-z}\left[1+\frac{1}{(n-1)!}\sum_{m=1}^{\infty}\frac{(-1)^m}{z^m}\frac{(2m-1)!}{2^{2m-1}(m-1)!}\sum_{k=0}^m\frac{(2k)!(n+m-k-1)!}{8^k(k!)^2(m-k)!}\right],\nonumber\\
\end{equation}

\noindent when $r\rightarrow\infty$. It is worth mentioning that Hem Prabha and Yadav \cite{HemPrabha96} proposed polynomial expressions for Bickley-Naylor fonctions up to $n$=7. Using Eq. (\ref{recur}), equation (\ref{g1}) can be put in the form

\begin{eqnarray}\label{g2}
g(z)=\left(\frac{7}{16}+\frac{z^2}{48}\right)K_0(z)-\left(\frac{9}{16}+\frac{z^2}{48}\right)Ki_1(z)-\left(\frac{19z}{48}+\frac{z^3}{48}\right)Ki_2(z)
\end{eqnarray}

\noindent and we therefore have to calculate 

\begin{eqnarray}\label{deltaV}
\delta V(r)&=&-\frac{2\alpha^2}{3r}\int_0^{\infty}dxx\rho(x)\left[\left(\frac{7}{16}+\frac{c^2\left(r+x\right)^2}{12}\right)K_0(2c(r+x))\right.\nonumber\\
& &-\left(\frac{9}{16}+\frac{c^2\left(r+x\right)^2}{12}\right)Ki_1(2c(r+x))-\left.\left(\frac{19c(r+x)}{24}+\frac{c^3(r+x)^3}{48}\right)Ki_2(2c(r+x))\right].\nonumber\\
\end{eqnarray}

We consider here the terms involving the argument $2c(r+x)$ in the exponential in Eq. (\ref{jo}), but the formalism can be applied in the same way to the part for which the argument of the exponential is $2c|r-x|$. If we keep the first expression of $g(z)$ given in Eq. (\ref{g1}), we have to consider three Bickley-Naylor functions, namely $Ki_1$, $Ki_3$ and $Ki_5$. The expression in Eq. (\ref{g2}) is simpler than Eq. (\ref{g1}), since it involves only Bickley functions $Ki_1$, $Ki_2$ and the usual Bessel function $K_0$. Another possibility would be to express $g(z)$ in terms of functions $K_0$, $K_1$ and $K_{i_1}$ \cite{Roesel77}:

\begin{eqnarray}
g(z)=\frac{\left(21+z^2+48\right)}{48}K_0(z)-\frac{\left(19z^2+z^4\right)}{48}K_1(z)-\frac{\left(27-18z^2-z^4\right)}{48}K_{i_1}(z).
\end{eqnarray}

\noindent In Eq. (\ref{deltaV}), we need to consider the six following functions: 

\begin{equation}\label{lesh}
\left\{\begin{array}{l}
H_1(x)=xK_0(2c(r+x)),\\ 
H_2(x)=x\left[2c(r+x)\right]^2K_0(2c(r+x)),\\ 
H_3(x)=xKi_1(2c(r+x)),\\ 
H_4(x)=x\left[2c(r+x)\right]^2Ki_1(2c(r+x)),\\ 
H_5(x)=x\left[2c(r+x)\right]Ki_2(2c(r+x)),\\
H_6(x)=x\left[2c(r+x)\right]^3Ki_2(2c(r+x)),
\end{array}\right.
\end{equation}

\noindent and the issue boils down to the calculation of

\begin{equation}
\frac{d^n}{dx^n}H_i(x),\;\;\;\; i=1, 6.
\end{equation}

Using Leibniz formula for the multiple derivative of a product, we obtain, for $H_6(x)$:

\begin{eqnarray}
\frac{d^n}{dx^n}\left\{x\left[2c(r+x)\right]^3Ki_2(2c(r+x))\right\}=\sum_{k=0}^n\bin{n}{k}\frac{d^k}{dx^k}Ki_2(2c(r+x)).\frac{d^{n-k}}{dx^{n-k}}\left\{x\left[2c(r+x)\right]^3\right\},\nonumber\\
\end{eqnarray}

\noindent where for $k\ge 0$:

\begin{equation}
\frac{d^{n+k}}{dz^{n+k}}Ki_n(z)=(-1)^n\frac{d^k}{dz^k}K_0(z)=(-1)^{n+k}K_k(z).
\end{equation}

\subsection{Calculation using power expansion}

One possibility to obtain an exact expression for the integral in Eq. (\ref{sum}) consists in expanding all the functions $H_i$, $i$=1, 6 in power series using the exact expression 

\begin{eqnarray}\label{kin}
Ki_n(z)&=&2^{n-2}\sum_{k=0}^{n-1}\frac{(-z/2)^k}{k!(n-k-1)!}\left[\Gamma\left(\frac{n-k}{2},0\right)\right]^2\nonumber\\
& &+(-z)^n\sum_{k=0}^ {\infty}\frac{(z/2)^k(2k)!}{(k!)^2(n+2k)!}\left[\vphantom{\ln\left(\frac{z}{2}\right)}\Phi(k+1)-\Phi(2k+1)+\Phi(2k+n+1)-\gamma_E-\ln\left(\frac{z}{2}\right)\right],\nonumber\\
\end{eqnarray}

\noindent where 

\begin{equation}
\Phi(k+1)=1+\frac{1}{2}+\frac{1}{3}+\cdots+\frac{1}{k}
\end{equation}

\noindent and then to use, after expanding the function $\ln(x)$ in power series, the expression

\begin{eqnarray}
\int_0^{\infty}\frac{y^k}{1+e^{\left[(y-\xi)/a\right]}}dy=\frac{\xi^{k+1}}{k+1}+\sum_{n=0}^{E\left(\frac{k-1}{2}\right)}(2n+1)\bin{k}{2n+1}a^{2n+2}\left(2-\frac{1}{2^{2n}}\right)\zeta(2n+2)\xi^{k-2n-1}+\mathcal{R},
\end{eqnarray}

\noindent with

\begin{eqnarray}\label{eqr}
\mathcal{R}=\sum_{n=1}^{\infty}(-1)^ne^{-n\xi/a}\int_0^{\infty}y^ke^{-ny/a}dy=k!\sum_{n=1}^{k}(-1)^n\frac{e^{-n\xi/a}}{n^{k+1}},
\end{eqnarray}

\noindent which is simpler than the expression published in Refs. \cite{Roesel77,Schucan65}:

\begin{eqnarray}
\int_0^{\infty}\frac{y^k}{1+e^{\left[(y-\xi)/a\right]}}dy&=&a^{k+1}\left\{\left(\frac{\xi}{a}\right)^{k+1}\frac{1}{k+1}-k!(-1)^kLi_{k+1}(-e^{-\xi/a})+2\pi\left(\frac{\xi}{a}\right)^{k}k!\right.\nonumber\\
& &\left.\times\sum_{p=0}^{\left[(k-1)/2\right]}\left(2^{2p+1}-1\right)\left(\frac{\pi a}{\xi}\right)^{2p+1}\frac{|B_{2p+2}|}{(k-1-2p)!(2p+1)!}\right\},\nonumber\\
\end{eqnarray}

\noindent involving Euler's polylogarithm

\begin{equation}\label{polylog}
Li_n(z)=\sum_{q=1}^{\infty}\frac{z^q}{q^n}.
\end{equation}

\noindent and Bernoulli number $B_p$. $\left[X\right]$ is the integer part of $X$.

\subsection{Calculation using recurrence relations}

It is interesting to point out that

\begin{equation}\label{inth}
\int_0^{\xi}H(y)dy,
\end{equation}

\noindent involved in Eq. (\ref{sum}) can be expressed through quantities of the kind

\begin{eqnarray}\label{gene}
I(p,q)=\int_{\gamma}^{\delta}y^pKi_q(y)dy,
\end{eqnarray}

\noindent where $p$ and $q$ are integers. The integrals $I(p,q)$ can then be calculated using Eq. (\ref{kin}). One can also remark that, integrating $I(p,q)$ by parts, we obtain the recurrence relation

\begin{eqnarray}
I(p,q)=\frac{1}{p+1}\left[\gamma^{p+1}Ki_q(\gamma)-\delta^{p+1}Ki_q(\gamma)-I(p+1,q-1)\right],
\end{eqnarray}

\noindent which can be initialized by

\begin{equation}
\left\{
\begin{array}{l}
I(0,0)=\int_{\gamma}^{\delta}K_0(y)dy=Ki_1(\gamma)-Ki_1(\delta)\\
I(1,0)=\int_{\gamma}^{\delta}yK_0(y)dy=\gamma K_1(\gamma)-\delta K_1(\delta).\\
\end{array}
\right.
\end{equation}

\noindent The integrals involved in the residual term $\mathcal{R}$ (see Eq. (\ref{eqr})) can be calculated in a similar manner, being expressed through quantities of the kind

%\begin{equation}
%\int_{\gamma}^{\delta}x^k\ln(x)dx=\frac{(\gamma^{k+1}-\delta^{k+1})}{(k+1)^2}-\frac{(\gamma^{k+1}\ln(\gamma)-\delta^{k+1}\ln(\delta))}{k+1}.
%\end{equation}The integrals involved in the residual term $R$ (see Eq. (\ref{eqr})) can be calculated in a similar manner. They can be expressed through quantities of the kind

\begin{eqnarray}\label{lnpq}
L_n(p,q)=\int_{\gamma}^{\delta}y^pe^{-\alpha n y}Ki_q(y)dy,
\end{eqnarray}

\noindent with $\alpha$ strictly positive. Integral (\ref{inth}) corresponds to the case $\alpha$=0. Integrating the right-hand side of Eq. (\ref{lnpq}) by parts, we find:

\begin{eqnarray}
L_n(p,q)=\frac{1}{\alpha n}\left[e^{-\alpha n\gamma}\gamma^pKi_q(\gamma)-e^{-\alpha n\delta}\delta^pKi_q(\delta)+pL_n(p-1,q)-L_n(p,q-1)\right],
\end{eqnarray}

\noindent which can be initialized by

\begin{equation}\label{init2}
\left\{
\begin{array}{l}
L_n(0,0)=\int_{\gamma}^{\delta}e^{-\alpha n y}K_0(y)dy\\
L_n(1,0)=\int_{\gamma}^{\delta}ye^{-\alpha n y}K_0(y)dy\\
L_n(0,1)=\int_{\gamma}^{\delta}e^{-\alpha n y}Ki_1(y)dy.
\end{array}
\right.
\end{equation}

Using the expression (see Ref. \cite{Abramowitz72}, 9.6.13 p. 375):

\begin{equation}
K_0(z)=\sum_{k=0}^{\infty}\frac{1}{(k!)^2}\left(\frac{z^2}{4}\right)^k\left[\psi(k+1)-\ln\left(\frac{z}{2}\right)\right],
\end{equation}

\noindent where

\begin{equation}
\psi(k+1)=-\gamma_E+\Phi(k+1),
\end{equation}

\noindent as well as Eq. (\ref{kin}) for $n$=1, we see that the evaluation of the integrals in Eq. (\ref{init2}) reduces to the calculation of simple integrals of the kind

\begin{equation}
\int_{\gamma}^{\delta}y^kdy=\frac{\delta^{k+1}-\gamma^{k+1}}{k+1}.
\end{equation}

\noindent In addition, the last integrals defined in Eq. (\ref{init2}) involve terms of the kind

\begin{eqnarray}
\int_{\gamma}^{\delta}y^ke^{-\alpha n y}dy=\frac{1}{(\alpha n)^{p+1}}\left[\Gamma(k+1,\gamma\alpha n)-\Gamma(k+1,\delta\alpha n)\right].
\end{eqnarray}

The approach presented above applies for all the functions $H_1$ to $H_6$ given in Eq. (\ref{lesh}), yielding an analytical expression for $\delta V(r)$. The quantity $\mathcal{R}$ is much smaller (a few orders of magnitude) than the main terms of the summation in Eq. (\ref{sum}), and the series in Eq. (\ref{sum}) requires 10 terms for a convergence of 1 \%.

\section{Asymptotic form of $g(z)$}\label{sec3}

The asymptotic expression of $g(z)$ for small values of $r$ can be determined following the method described in Ref. \cite{Berestetskii82} for the Uehling potential of a point-like nucleus. We first split the integral in two parts:

\begin{eqnarray}
\int_1^{\infty}\frac{\sqrt{t^2-1}}{t^3}\left(1+\frac{1}{2t^2}\right)e^{-2ctr}dt=\int_1^{\infty}\frac{\sqrt{t^2-1}}{t^3}e^{-2ctr}dt+\int_1^{\infty}\frac{\sqrt{t^2-1}}{2t^5}e^{-2ctr}dt=I_1+I_2
\end{eqnarray}

\noindent and choose $t_1$ ($\frac{1}{cr}\gg t_1\gg 1$), such that

\begin{eqnarray}
I_1=\int_1^{t_1}\frac{\sqrt{t^2-1}}{t^3}e^{-2ctr}dt+\int_{t_1}^{\infty}\frac{\sqrt{t^2-1}}{t^3}e^{-2ctr}dt=J_1+J_2.
\end{eqnarray}

\noindent $J_1$ can be estimated setting $r=0$  and the change of variable $u^2=t^2-1$ yields

\begin{eqnarray}
J_1=\int_0^{\sqrt{t_1^2-1}}\frac{u^2}{\left(u^2+1\right)^2}du=\frac{1}{2}\left(-\frac{\sqrt{t_1^2-1}}{t_1^2}+\arctan\left(\frac{1}{\sqrt{t_1^2-1}}\right)\right).\nonumber\\
\end{eqnarray}

\noindent For $t_1\rightarrow\infty$, we get $J_1\rightarrow\pi/4$. The other integrals are easy to evaluate. In $J_2$, we can neglect 1 in the square root, which yields, after two successive integrations by parts

\begin{eqnarray}
J_2=\int_{t_1}^{\infty}\frac{1}{t^2}e^{-2ctr}dt=\frac{1}{t_1}e^{-2ct_1r}+2cr\ln(t_1)~e^{-2ct_1r}-(2cr)^2\int_{t_1}^{\infty}\ln(t)~e^{-2ctr}dt.
\end{eqnarray}

\noindent The first two terms tend to zero when $t_1\rightarrow\infty$ and the last term is

\begin{equation}
\int_{t_1}^{\infty}\ln(t)~e^{-2ctr}dt=\int_{2ct_1r}^{\infty}\frac{\left[\ln(u)-\ln(2cr)\right]}{2cr}~e^{-u}du
\end{equation}

\noindent yielding, for $r$ close to zero

\begin{equation}
\int_{0}^{\infty}\ln(t)~e^{-2ctr}dt=\frac{1}{2cr}\left[\ln\left(\frac{1}{2cr}\right)-\gamma_E\right]
\end{equation}

\noindent and therefore, for $r\rightarrow 0$, we have

\begin{equation}
J_2\approx-2cr\ln\left(\frac{1}{2cr}\right).
\end{equation}

\noindent For $I_2$ we can set directly $r=0$:

\begin{equation}
I_2=\int_1^{\infty}\frac{\sqrt{t^2-1}}{2t^5}dt=\frac{\pi}{32}
\end{equation}

\noindent and then

\begin{equation}
I_1+I_2=\frac{\pi}{4}+\frac{\pi}{32}-2cr\ln\left(\frac{1}{2cr}\right),
\end{equation}

\noindent leading to the asymptotic form:

\begin{equation}
g(z)\approx\frac{9\pi}{32}-2cr\ln\left(\frac{1}{2cr}\right).
\end{equation}

\noindent This result can be obtained by integrating the quantity $-3\pi r\delta V(r)/\left(2\alpha Z\right)$ in expression (\ref{as1}) with respect to variable $(2cr)$, noticing that

\begin{equation}
g(0)=\int_1^{\infty}\frac{\sqrt{t^2-1}}{2t^5}\left(2t^2+1\right)dt=\frac{9\pi}{32}.
\end{equation}

For large values of $r$ we find the same asymptotic form

\begin{equation}
\delta V(r)\approx-\frac{Z}{r}\left(1+\frac{\alpha}{4\sqrt{\pi}}\frac{e^{-2cr}}{\left(cr\right)^{3/2}}\right),
\end{equation}

\noindent which is the same as for the Uehling potential in the pure Coulomb case.

\section{Higher-order contribution: the K\"all\`en-Sabry potential}\label{sec4}

The procedure presented for the Uehling potential can be used for the calculation of fourth-order QED corrections in $\alpha^2(Z\alpha)$ (the corresponding Feynman diagrams are the two-loop diagram and three diagrams with an additional photon line within a single electron-positron loop) using the K\"all\`en-Sabry potential \cite{Kallen55,Schneider93,Indelicato13}:

\begin{eqnarray}
V_{KS}(r)=\frac{\alpha^2\left(Z\alpha\right)}{\pi^2 r}\int_0^{\infty}dxx\rho(x)\left[L_0(2c|r-x|)-L_0(2c(r+x))\right],
\end{eqnarray}

\noindent where

\begin{equation}
L_0(x)=-\int^ xL_1(u)du,
\end{equation}

\noindent with

\begin{eqnarray}
L_1(u)&=&\int_1^{\infty}\left\{\left(\frac{2}{3t^5}-\frac{8}{3t}\right)f(t)+\left(\frac{2}{3t^4}+\frac{4}{3t^2}\right)\sqrt{t^2-1}\ln\left[8t(t^2-1)\right]\right\}\nonumber\\
& &+\sqrt{t^2-1}\left(\frac{2}{9t^6}+\frac{7}{108t^4}+\frac{13}{54t^2}\right)+\left(\frac{2}{9t^7}+\frac{5}{4t^5}+\frac{2}{3t^3}-\frac{44}{9t}\right)\ln\left[\sqrt{t^2-1}+t\right]e^{-ut}dt\nonumber\\
\end{eqnarray}

\noindent and

\begin{eqnarray}\label{fonctionf}
f(t)=\int_1^{\infty}\left[\frac{(3x^2-1)\ln\left[\sqrt{x^2-1}+x\right]}{x(x^2-1)}-\frac{\ln\left[8x(x^2-1)\right]}{\sqrt{x^2-1}}\right]dx.
\end{eqnarray}

\noindent An exact expression of $f(t)$ is given in Appendix A. The method proposed in the present paper for the Uehling potential can be applied to the K\"all\`en-Sabry potential using the fit proposed by Indelicato \cite{Indelicato13}:

\begin{equation}
L_1(u)=\left(a+b\sqrt{u}+cu+du^{3/2}+eu^2+fu^{5/2}\right)\frac{e^{-u}}{u^{7/2}}
\end{equation}

\noindent for $u>3$. $L_0(x)$ is then obtained by direct integration of the latter expression, the integration constant being fixed assuming that $L_0\rightarrow 0$ when $r\rightarrow \infty$. For $u\le 3$, the form, inspired from Blomqvist \cite{Blomqvist72}, is

\begin{equation}
L_1(u)=uh_2(u)\left[\ln(u)\right]^2+uh_1(u)\ln(u)+h_0(u).
\end{equation}

\noindent The coefficients $a$, $b$, $c$, $d$, $e$ and $f$ are given in Appendix A of Ref. \cite{Indelicato13} and functions $h_0(u)$, $h_1(u)$ and $h_2(u)$, also derived by Indelicato, in Appendix B of the latter article (there are two small typos in Appendix B of Ref. \cite{Indelicato13}: $g_1$ and $g_2$ should be replaced by $h_1$ and $h_2$ respectively).

\section{Conclusion}

We proposed a closed formula for the Uehling potential in case of a Fermi charge distribution. The result combines a Sommerfeld-type expansion of the involved integral together with Bickley-Naylor functions and Bessel functions of the second kind. The Fermi distribution is widely used in QED computations in order to overcome the assumption of a point-like nucleus. The obtained expression is an extension of the result obtained by Frolov and Wardlaw in the pure Coulomb case and enables one to avoid numerical integration and analytical fitting formulas. The relations given here may also serve as guides for the derivation of rational approximations. We do not pretend that the formulas presented in this work are likely to bring any significant improvement in numerical accuracy or speed. The most efficient method to compute the Uehling potential for any nuclear charge distribution (and in particular for the Fermi or Woods-Saxon distribution) is probably the rational approximation published by Fullerton and Rinker, which provides nine-digit accuracy with a low numerical cost. Thus, the most natural solution is to use this rational approximation, and to evaluate the integral by usual numerical integration methods. We would be happy if the mathematical expressions, relations and properties discussed in the present article could help to bring new ideas in the field. In the future, we plan to investigate fourth-order QED corrections in $\alpha^2(Z\alpha)$ using the K\"all\`en-Sabry potential \cite{Kallen55,Schneider93,Indelicato13}, still in the case of a Fermi nuclear charge distribution.

%, and orders $\alpha(Z\alpha)^3$, $\alpha(Z\alpha)^5$ and $\alpha(Z\alpha)^7$, which represent the major effect of the distortion of the electron and positron wavefunction in a strong Coulomb field \cite{Wichmann56,Blomqvist72,Fricke69}.   

% apologies : not numerical
% Mezo also provided an analytical expression
% asymptotic properties : nw
% Kallen-Sabry: new
% expression exclusively in terms of Bickley functions : new. 

\appendix

\section{Analytical expression for the function $f(t)$ involved in the K\"all\`en-Sabry potential} 

The function $f(t)$, defined in Eq. (\ref{fonctionf}), is equal to

\begin{eqnarray}
f(t)=\frac{2\pi^2}{3}-\ln(\eta)\ln\left[\frac{\left(\eta^ 4-1\right)\left(\eta^ 2-1\right)}{\eta^ 2}\right]+Li_2\left(-\frac{1}{\eta^2}\right)-2\Re\left[Li_2\left(\eta^2\right)\right],
\end{eqnarray}

\noindent where $\eta=t+\sqrt{t^2-1}$, $Li_2(z)$ is the dilogarithm function defined in Eq. (\ref{polylog}) and $\Re$ the real part. The integral form of the dilogarithm function is

\begin{equation}
Li_2(z)=-\int_0^1\frac{\ln(1-zt)}{t}dt.
\end{equation}

%Author, \textit{Book title} (Publisher, place year) page numbers
% etc

\end{document}